\documentclass[preprint,superscriptaddress,aps]{revtex4}
\usepackage{amsfonts}
\usepackage{amssymb}
\usepackage{amsmath}
\usepackage{graphics}
\usepackage{graphicx}
\usepackage{color}
\usepackage[T1]{fontenc}
\usepackage{hyperref}
\newcommand{\bea}{\begin{eqnarray}}
\newcommand{\eea}{\end{eqnarray}}

\begin{document}
\title{On the effective superpotential in the supersymmetric 
Chern-Simons theory with matter}

\author{M. Gomes}
\email{mgomes@fma.if.usp.br}
\affiliation{Instituto de F\'\i sica, Universidade de S\~ao Paulo\\
Caixa Postal 66318, 05315-970, S\~ao Paulo, SP, Brazil}

\author{A.~C.~Lehum}
\email{andrelehum@ect.ufrn.br}
\affiliation{Escola de Ci\^encias e Tecnologia, Universidade Federal
  do Rio 
Grande do Norte\\
Caixa Postal 1524, 59072-970, Natal, RN, Brazil}

\author{J. R. Nascimento}
\email{jroberto@fisica.ufpb.br}
\affiliation{Departamento de F\'{\i}sica, Universidade Federal da Para\'{\i}ba\\
 Caixa Postal 5008, 58051-970, Jo\~ao Pessoa, Para\'{\i}ba, Brazil}

\author{A. Yu. Petrov}
\email{petrov@fisica.ufpb.br}
\affiliation{Departamento de F\'{\i}sica, Universidade Federal da Para\'{\i}ba\\
 Caixa Postal 5008, 58051-970, Jo\~ao Pessoa, Para\'{\i}ba, Brazil}

\author{A. J. da Silva}
\email{ajsilva@fma.if.usp.br}
\affiliation{Instituto de F\'\i sica, Universidade de S\~ao Paulo\\
Caixa Postal 66318, 05315-970, S\~ao Paulo, SP, Brazil}

\begin{abstract}
We develop a superfield approach to the effective potential in the
supersymmetric Chern-Simons theory coupled to matter and, in the Landau gauge,
calculate the one-loop K\"{a}hlerian effective potential and give some
qualitative prescriptions for the twe-loop one.
\end{abstract}

\maketitle

It is well known that the low-energy effective dynamics of a quantum
field theory is characterized by the effective
potential~\cite{CW}. Its study is especially relevant in the context
of supersymmetric theories since the effective (super)potential allows
to obtain important information about supersymmetry and/or gauge
symmetry breaking. In four space-time dimensions, the methodology of
studying the effective superpotential has been well developed in
\cite{BK0,WZ} and then successfully applied in many examples,
including the Wess-Zumino model, both in commutative~\cite{WZ} and
noncommutative~\cite{WZNC} cases, general chiral superfield
model~\cite{GC}, super-Yang-Mills theory~\cite{SYM} and
higher-derivative theories \cite{HD}. 

Recently, a great deal of attention has been devoted to studies of the
supersymmetric Chern-Simons theory with matter. The reason is the fact
that $N=6$ and $N=8$ supersymmetric Chern-Simons theories are finite
and conformal invariant, which allows for studying
the $AdS_4/CFT_3$ correspondence \cite{Mald}. Earlier, different
aspects of the supersymmetric Chern-Simons theories have been studied
in  \cite{CS}. 

In three space-time dimensions,
the first attempt to formulate a superfield approach to the study of
the supersymmetric effective potential  has been carried out in
\cite{ourEP,ourEP1} (some preliminary discussions of this approach were
presented in \cite{Burgess,Lehum}, see also \cite{BSM} for some issues
related to extended supersymmetry in these  theories).
Additional studies of the effective potential including
a component analysis of the Coleman-Weinberg and Wess-Zumino models
were performed in \cite{Gall}. 

In this work we consider the superfield version of the 
supersymmetric Coleman-Weinberg model, that is, the Chern-Simons
theory coupled to a self-interacting massless scalar matter,
which is the simplest example of a three-dimensional supersymmetric
gauge theory with matter, where all couplings are
dimensionless. Employing an adequate method, by us devised, we
will calculate the one- and discuss the two-loop superpotentials. 
Throughout the
paper, we follow the notations and conventions adopted in  \cite{SGRS}.

The action of the Coleman-Weinberg theory looks like
\begin{eqnarray}
\label{sfa}
S[\Phi,A^{\alpha}]&=&\int d^5z
\left[A^{\alpha}W_{\alpha}-\frac{1}{2}(D^{\alpha}-
igA^{\alpha})\Phi
(D_{\alpha}+igA_{\alpha})\bar{\Phi}+V(\Phi)+\right.\nonumber\\&+&\left.
\frac{1}{2\xi}\int
d^5z (D^{\alpha}A_{\alpha}) (D^{\beta}A_{\beta})\right],
\end{eqnarray}
where $\Phi$ is a (complex) scalar superfield and $V(\Phi)=
\frac{\lambda}{2}(\Phi\bar{\Phi})^2$ is a
classical potential.
The $A_{\alpha}$ is a gauge superfield, with 
$W_{\alpha}=\frac{1}{2}D^{\beta}D_{\alpha}A_{\beta}$ the
corresponding gauge-invariant superfield strength.
The last term of the expression (\ref{sfa}) is the gauge-fixing action.
Since the theory is Abelian, the ghosts completely decouple, and their
action 
will be omitted.

We will start by evaluating the superfield effective action, within
the methodology of the loop expansion~\cite{BO}. To do it, we make a
shift $\Phi\to\Phi+ \, \phi$ in the superfield $\Phi$  
(the shift for the $\bar{\Phi}$ will be similar),
where now $\Phi$ is a background (super)field, and $\phi$ is a quantum
one. 
Through this study, the gauge field $A_{\alpha}$ is taken
to be purely quantum since our aim consists in studying quantum
corrections depending only on the scalar fields. As a result, the
classical action (\ref{sfa}) takes the form
\begin{eqnarray}
\label{2}
S[\Phi;\phi,A^{\alpha}]&=&S[\Phi,A^{\alpha}]|_{A^{\alpha}=0}+
\int d^5z[\frac{1}{2}A_{\alpha}
(D^{\beta}D^{\alpha}+\frac{1}{\xi} D^{\alpha} D^{\beta})A_{\beta}+
\phi D^2\bar{\phi}+\nonumber\\&+&
\frac{\lambda}{2}(\Phi^2\bar{\phi}^2+4\Phi\bar{\Phi}\phi\bar{\phi}+\bar{\Phi}^2
\phi^2)+\frac{ig}{2}(\Phi
A^{\alpha}D_{\alpha}\bar{\phi}-\bar{\Phi}A^{\alpha}D_{\alpha}\phi)+\nonumber\\&+&
\frac{ig}{2}(\phi
A^{\alpha}D_{\alpha}\bar{\Phi}-
\bar{\phi}A^{\alpha}D_{\alpha}\Phi)
+\lambda(\phi^2\bar{\phi}\bar{\Phi}+\bar{\phi}^2\phi\Phi+ 
\frac{1}{2}(\phi\bar{\phi})^2)+
\nonumber\\&+&\frac{ig}{2}(\phi
A^{\alpha}D_{\alpha}\bar{\phi}-\bar{\phi}A^{\alpha}D_{\alpha}\phi)-
\frac{g^2}{2}A^{\alpha}A_{\alpha}(\Phi\bar{\Phi}+\Phi\bar{\phi}+\phi\bar{\Phi}+
\phi\bar{\phi})].
\end{eqnarray}
Here we eliminated the linear terms in quantum fields since they
produce only irrelevant, one-particle-reducible  contributions.
The effective action $\Gamma[ \Phi]$ is defined by the expression
\begin{eqnarray}
\exp\left(i\Gamma[\Phi]\right)\,= \,\int D\phi
DA^{\alpha} \, 
\exp\left(iS[\Phi,\phi,A^{\alpha}]\right)\Big|_{1PI},
\end{eqnarray}
where the subscript $1PI$ stands for one-particle-irreducible supergraphs.
The general structure of the effective action can be cast in a form
similar to the four-dimensional case~\cite{BK0,WZ}:
\begin{eqnarray}
\Gamma[\Phi]=\int d^5z \, K(\Phi)+\int d^5z \, F(D^{\alpha}\Phi D_{\alpha}\Phi,D^2\Phi;\Phi),
\end{eqnarray}
where the $K(\Phi)$ is the K\"{a}hlerian effective potential which
depends only on the superfield $\Phi$ but not on its derivatives. The
$F$ term is called auxiliary fields' effective potential whose key
property is its vanishing in the case when all derivatives of the
superfields are equal to zero (within this paper we will not
discuss it). We restrict ourselves to the K\"{a}hlerian effective potential.

We will work within a loop expansion for the effective action $\Gamma$, 
\begin{eqnarray}
\label{4}
\Gamma[ \Phi]=S[ \Phi]+ \Gamma^{(1)}[ \Phi]+\Gamma^{(2)}[
\Phi]+
\ldots,
\end{eqnarray}
and for the K\"ahlerian potential $K$,
\begin{eqnarray}
K(\Phi)=V(\Phi)+\sum_{L=1}^{\infty}K_L(\Phi), 
\end{eqnarray}
that is, the tree-order K\"{a}hlerian effective potential is 
$K^{(0)}(\Phi)=V(\Phi)=\frac{\lambda}{2}(\Phi\bar{\Phi})^2$.

For the background fields equal to zero, the free propagators of the
scalar and gauge  superfields corresponding to the action (\ref{sfa}) are
\bea
<A^{\alpha}(z_1)A^{\beta}(z_2)>&=&\frac{i}{4\Box}(D^{\beta}D^{\alpha}+
\xi D^{\alpha}D^{\beta})\delta^5(z_1-z_2);\nonumber\\
<\phi(z_1)\bar{\phi}(z_2)>&=&-\frac{i}{\Box}D^2\delta^5(z_1-z_2).
\eea

The key point of this paper consists in the development of a methodology for the background dependent propagators and their use for the calculation of the effective action. While in the four-dimensional superfield theories this formalism has been well developed and successfully applied in a number of papers, see f.e. \cite{WZ,GC,SYM,HD}, in the three-dimensional superfield theories it has been used only in \cite{ourEP} for the purely scalar superfield model. Also, in the works \cite{ourEP1} a simplified form of this methodology based on the  imposition of restrictions on the structure of the background scalar field ($\Phi=\phi_1-\theta^2\phi_2$) was  used, whereas in the present paper the general form $\Phi=\phi_1+\theta^{\alpha}\psi_{\alpha}-\theta^2\phi_2$ is used, with no restrictions except for the condition $D_{\alpha}\Phi=0$, leading to the K\"{a}hlerian effective potential; actually, we do not employ the component expansion of $\Phi$. Moreover, we generalize the powerful technique of summation over one-loop diagrams (which has been intensively applied in the four dimensions \cite{GC,HD}) for the three-dimensional case, specially, for the supersymmetric gauge theories, which is done here for the first time.  We expect that this methodology can be efficiently applied to more sophisticated theories. 

It will be convenient to obtain the effective propagators on the base
of summation of different sequences of free propagators. First of
all, we can take into account the situation with the triple
gauge-scalar vertices. Including such a vertex into a diagram, with
one background scalar leg, produces a fragment of the
diagram as shown in Fig.1.

\begin{figure}[ht]
\centerline{\includegraphics{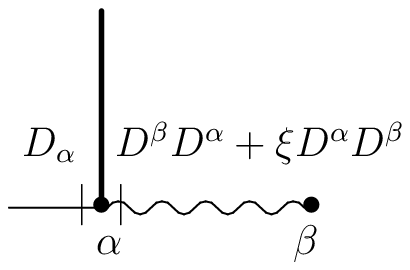}} 
\caption{A fragment of a Feynman diagram involving triple vertices.}
\end{figure}

In that figure the bold line is for the background field, and the factor
$D^{\beta}D^{\alpha}+\xi D^{\alpha}D^{\beta}$ originates from the
gauge propagator. The indices $\alpha$ and $\beta$ are the indices of
gauge fields contracted into this propagator. We see that if we move
the derivative $D_{\alpha}$, originated from the interaction vertex,
to the propagator of the gauge fields proportional to
$D^{\beta}D^{\alpha}+\xi D^{\alpha}D^{\beta}$, we will annihilate its
gauge-independent part, and the gauge-dependent part is proportional to
$\xi$, by imposing the gauge $\xi=0$ (Landau gauge
\cite{SGRS}), the calculations are simplified, removing from
considerations all diagrams with the vertices $(\Phi
A^{\alpha}D_{\alpha}\bar{\phi}-\bar{\Phi}A^{\alpha}D_{\alpha}\phi)$.

Let us find the one-loop K\"{a}hlerian effective
potential. Taking into account the conclusion made in the previous
paragraph, we can find that in the Landau gauge there are two types of
contributions to the K\"{a}hlerian effective potential: in the first
of them, all supergraphs involve only the gauge propagators, and in
the second of them, all supergraphs involve only matter
propagators. It is equivalent to write down two contributions
to the one-loop effective action as two traces of the logarithms:
\bea
\Gamma^{(1)}_1&=&\frac{i}{2}{\rm Tr}\ln\Big[D^{\beta}D^{\alpha}+\frac{1}{\xi}
D^{\alpha} 
D^{\beta}+g^2C^{\alpha\beta}\Phi\bar{\Phi}
\Big];\nonumber\\
\Gamma^{(1)}_2&=&-\frac{i}{2}{\rm Tr}\ln\left(\begin{array}{cc} 
M & D^2+m\\
D^2+m & \bar{M}
\end{array}\right),
\eea 
where $m=2\lambda\Phi\bar{\Phi}$, $\bar{M}=\lambda\Phi^2$, 
$M=\lambda\bar{\Phi}^2$. The plus sign in $\Gamma^{(1)}_1$ arose due to the
fermionic statistics of $A_{\alpha}$.
The $\Gamma_1^{(1)}$, at the $\xi=0$ limit, can be explicitly obtained
via the expansion of the trace of the logarithm: since the operator
inverse to ${\cal O}^{\alpha\beta}=D^{\beta}D^{\alpha}+\frac{1}{\xi}
D^{\alpha} D^{\beta}$ is
$G_{\beta\gamma}=\frac{i}{4\Box}(D_{\gamma}D_{\beta}+\xi
D_{\beta}D_{\gamma})\delta^5(z-z')$ (i.e. ${\cal
  O}^{\alpha\beta}G_{\beta\gamma}=-i\delta^{\alpha}_{\gamma}\delta^5(z-z')$), we can
write $\Gamma^{(1)}_1$, at $\xi=0$,  as
\bea
\Gamma^{(1)}_1=\frac{i}{2}{\rm
  Tr}\sum_{n=1}^{\infty}\frac{1}{n}(\frac{g^2\Phi\bar{\Phi}}{4\Box})^n
D^{\alpha_2}D_{\alpha_1}D^{\alpha_3}D_{\alpha_2}\ldots
D^{\alpha_1}
D_{\alpha_n}\delta_{12}|_{\theta_1=\theta_2}.
\eea
Here we took into account that each relevant scalar-gauge vertex looks
like $-\frac{1}{2}g^2\Phi\bar{\Phi}A^{\alpha}A_{\alpha}$, and  
$<A_{\alpha}(z_1)A^{\beta}(z_2)>=\frac{i}{4\Box}D^{\beta}D_{\alpha}\delta_{12}$.
Using the commutator for the supercovariant derivatives 
$[D^{\gamma},D_{\beta}]=2\delta^{\gamma}_{\beta}D^2$ together with the
property $D^{\alpha}D_{\beta}D_{\alpha}=0$, one finds that
$D^{\alpha_2}D_{\alpha_1}D^{\alpha_3}D_{\alpha_2}\ldots 
D^{\alpha_1}D_{\alpha_n}\delta_{12}|_{\theta_1=\theta_2}=2^n\Box^{(n-1)/2}$,
for 
$n=2l+1$; instead, if $n=2l$ is even, this expression vanishes. Hence we have
\bea
\Gamma^{(1)}_1=\frac{i}{2}\int d^3x_1d^2\theta
\sum_{l=0}^{\infty}\frac{1}{2l+1}
\Big(\frac{g^2\Phi\bar{\Phi}}{4\Box}\Big)^{2l+1}2^{2l+1}\Box^l
\delta^3(x_1-x_2)|_{x_1=x_2}.
\eea
Then, by carrying out the Fourier transform ($\Box\to -k^2$) we arrive at
the following contribution to the K\"{a}hlerian effective potential
(as usual, the corresponding effective action can be restored from the
relation $\Gamma^{(1)}_1=\int d^5z K^{(1)}_1$):
\bea
K^{(1)}_1=-i\int\frac{d^3k}{(2\pi)^3}
\sum_{l=0}^{\infty}\frac{(-1)^l}{2l+1}(\frac{\mu^2}{2k^2})^{2l+1}(k^2)^l,
\eea
where $\mu^2=g^2\Phi\bar{\Phi}$.
To do the summation, we can consider 
\bea
\label{dif}
\frac{dK^{(1)}_1}{d\mu^2}=-\frac{i}{2}\int\frac{d^3k}{(2\pi)^3k^2}
\sum_{l=0}^{\infty}\frac{(-1)^l}{2^{2l+1}}(\frac{\mu^2}{\sqrt{k^2}})^{2l}.
\eea
This expression can be rearranged, summed and integrated with use of
the dimensional regularization as 
\bea
&&\frac{dK^{(1)}_1}{d\mu^2}=\frac{1}{4}\int\frac{d^3k}{(2\pi)^3k^2}
\sum_{l=0}^{\infty}(-\frac{\mu^4}{4k^2})^l=
\frac{1}{4}
\int\frac{d^3k}{(2\pi)^3}
\frac{1}{k^2+\frac{1}{4}\mu^4}=-\frac{\mu^2}{32\pi}.
\eea
Here we performed a Wick rotation which yields $k^2_E=k^2$.
All this allows us to integrate the equation (\ref{dif}) for $K^{(1)}_1$ 
and write down the following
contribution 
to the one-loop effective action
\bea
\label{k1}
\Gamma^{(1)}_1=-\int d^5 z\frac{(g^2\Phi\bar{\Phi})^2}{64\pi}.
\eea
This expression, in {\bf its functional structure}, is similar to the results
of \cite{ourEP}. Indeed, it is finite, polynomial and
does not involve any logarithm-like  dependence.

At the same time, we can find the contribution from the purely matter sector:
\bea
\Gamma^{(1)}_2&=&-\frac{i}{2}{\rm Tr}\ln\left( D^2 \left(\begin{array}{cc} 
0 & 1\\
1 & 0
\end{array}\right)+{\cal M}\right),
\eea
where ${\cal M}=\left(\begin{array}{cc} 
M & m\\
m & \bar{M}
\end{array}\right)$. One can elaborate this expression via expansion
in power 
series, which yields
\bea
\Gamma^{(1)}_2&=&\frac{1}{2}\sum_{n=0}^{\infty}\frac{1}{2n+1}{\rm Tr}
\int\frac{d^3k_E}{(2\pi)^3}\frac{\tilde{\cal M}^{2n+1}}{(k^2)^{n+1}},
\eea
where $\tilde{\cal M}={\cal M}\left(\begin{array}{cc} 
0 & 1\\
1 & 0
\end{array}\right)=\left(\begin{array}{cc} 
m & M\\
\bar{M} & m
\end{array}\right)$.
The explicit result can be obtained via the diagonalization of the
matrix 
$\tilde{\cal M}$. As the matrix is Hermitian, its diagonal form can be
obtained by  calculating its eigenvalues which are 
$\lambda_{1,2}=m\pm\sqrt{M\bar{M}}$. 
This means that
\bea
\Gamma^{(1)}_2&=&\frac{1}{2}\sum_{n=0}^{\infty}\frac{1}{2n+1}
\int\frac{d^3k}{(2\pi)^3}\frac{\lambda_1^{2n+1}+\lambda_2^{2n+1}}{(k^2)^{n+1}}.
\eea
This sum can be evaluated in the same way as above yielding
\bea
\label{k2}
\Gamma^{(1)}_2=\frac{1}{8\pi}\int d^5z
(\lambda^2_1+\lambda^2_2)=\frac{1}{4\pi}
\int d^5z(M\bar{M}+m^2).
\eea
Hence, the complete one-loop K\"{a}hlerian effective potential can be
read off from the sum of  (\ref{k1}) and (\ref{k2}):
\bea
\label{1loop}
K^{(1)}=-\frac{(g^2\Phi\bar{\Phi})^2}{64\pi}+
\frac{5\lambda^2(\Phi\bar{\Phi})^2}{8\pi}.
\eea
As it was already noted, it is finite and polynomial.

Now, let us turn to the two-loop approximation which we restrict
to some qualitative remarks.

First we obtain the effective propagator of  
$A_{\alpha}$ field via the sum depicted at Fig. 2.

\begin{figure}[ht]
\centerline{\includegraphics{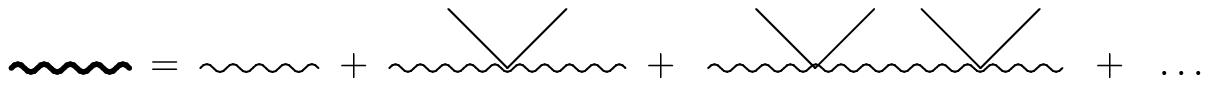}} 
\caption{Calculation of the ``dressed'' propagator.}
\end{figure}

To do it, we start with the quadratic action of the $A_{\alpha}$ field:
\bea
\label{2a}
S_2=\frac{1}{2}\int
d^5zA_{\alpha}\Big[(D^{\beta}D^{\alpha}+\frac{1}{\xi} D^{\alpha}  
D^{\beta})+\mu^2C^{\alpha\beta}
\Big]A_{\beta},
\eea
which can be read off from (\ref{2}). As we already noted, after the
calculation,  we must impose $\xi=0$.
Then, we employ the identity:
\bea
C^{\alpha\beta}=-\frac{1}{2D^2}(D^{\alpha}D^{\beta}-D^{\beta}D^{\alpha}),
\eea
which straightforwardly follows from the well-known relation
$D^{\alpha}D^{\beta}=i\partial^{\alpha\beta}-C^{\alpha\beta}D^2$
(which implies $[D^{\alpha},D^{\beta}]=-2C^{\alpha\beta}D^2$).
To obtain the background-dependent propagator from (\ref{2a}), we must
invert the  operator
\bea
\label{oper}
{\cal O}^{\alpha\beta}=D^{\beta}D^{\alpha}+\frac{1}{\xi} D^{\alpha}
D^{\beta}+ C^{\alpha\beta}\mu^2=
D^{\beta}D^{\alpha}(1+\frac{\mu^2}{2D^2}) +D^{\alpha}
D^{\beta}(\frac{1}{\xi}- \frac{\mu^2}{2D^2}).
\eea
It is easy to verify that if ${\cal O}^{\alpha\beta}=
A D^{\beta}D^{\alpha}+B D^{\alpha} D^{\beta}$, then the corresponding
propagator, 
that is, $G_{\beta\gamma}$ defined such that ${\cal
  O}^{\alpha\beta}G_{\beta\gamma}
=-i\delta^{\alpha}_{\gamma}\delta^5(z-z')$, looks like
\bea
G_{\beta\gamma}=(G_1 D_{\beta}D_{\gamma}+G_2D_{\gamma}D_{\beta})\delta^5(z-z'),
\eea
where
\bea
G_1=-\frac{i}{4B\Box};\quad\, G_2=-\frac{i}{4A\Box}.
\eea

Replacing the values $A=1+\frac{\mu^2}{2D^2}$ and 
$B=\frac{1}{\xi}-\frac{\mu^2}{2D^2}$, see (\ref{oper}), we find
\bea
G_{\beta\gamma}(z_1,z_2)=\frac{i}{4\Box}(\frac{\xi
  D_{\beta}D_{\gamma}}{1-\frac{\xi
    M^2}{2D^2}}+\frac{D_{\gamma}D_{\beta}}{1+\frac{\mu^2}{2D^2}})
\delta^5(z_1-z_2).
\eea
Imposing the Landau gauge $\xi=0$, we get
\bea
\label{vecpro}
<A^{\alpha}(z_1)A^{\beta}(z_2)>&=&\frac{i}{4\Box+2\mu^2D^2}D^{\beta}D^{\alpha}
\delta^5(z_1-z_2).
\eea

The next step consists in obtaining the background dependent
propagator for the superfields $\phi,\bar{\phi}$. To do it, we consider the
quadratic action of only these quantum fields. As we have already
noted before, it looks like
\bea
S_2[\phi,\bar{\phi}]=\frac{1}{2}\int d^5z 
\left(\begin{array}{cc}\phi & \bar{\phi}
\end{array}\right)\left(\begin{array}{cc} 
M & D^2+m\\
D^2+m & \bar{M}
\end{array}\right)\left(\begin{array}{c}\phi \\ \bar{\phi}
\end{array}\right).
\eea
Therefore, the matrix propagator is
\bea
\label{scapro}
&&\left(\begin{array}{cc} 
<\phi(z_1)\phi(z_2)> & <\phi(z_1)\bar{\phi}(z_2)>\\
<\bar{\phi}(z_1)\phi(z_2)> & <\bar{\phi}(z_1)\bar{\phi}(z_2)>
\end{array}\right)=\nonumber\\
&=&-i\left(\begin{array}{cc} 
\bar{M} & -(D^2+m)\\
-(D^2+m) & M
\end{array}\right)\frac{1}{M\bar{M}-(D^2+m)^2}\delta^5(z_1-z_2).
\eea
Thus, we have found all background-dependent propagators.

The interaction part of the action (\ref{2}), after absorbing some
terms into the propagator and removing irrelevant terms mentioned
above looks like
\begin{eqnarray}
\label{2int}
S_{int}[\Phi;\phi,A^{\alpha}]&=&\int d^5z[
\lambda(\phi^2\bar{\phi}\bar{\Phi}+\bar{\phi}^2\phi\Phi+ 
\frac{1}{2}(\phi\bar{\phi})^2)+
\nonumber\\&+&\frac{ig}{2}(\phi
A^{\alpha}D_{\alpha}\bar{\phi}-\bar{\phi}A^{\alpha}D_{\alpha}\phi)+
g^2A^{\alpha}A_{\alpha}(\Phi\bar{\phi}+\phi\bar{\Phi}+
\phi\bar{\phi})].
\end{eqnarray}

\begin{figure}[ht]
\centerline{\includegraphics{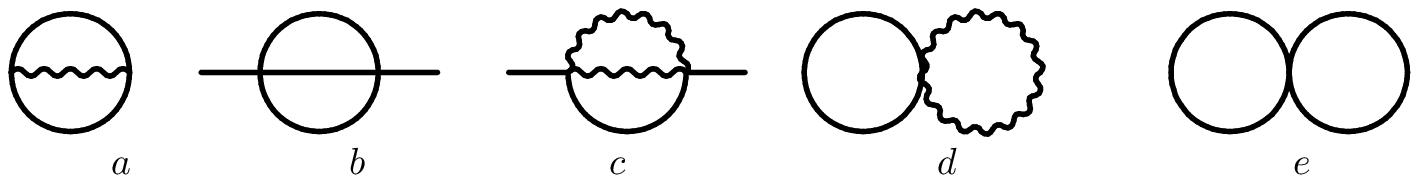}} 
\caption{The two-loop Feynman supergraphs.}
\end{figure}

Using the vertices from this expression, we must calculate the
contributions from the diagrams depicted in Fig. 3. We will
not calculate these diagrams exactly, but the dimensional analysis and
a straightforward inspection of their contributions show
that the two-loop K\"{a}hlerian effective potential has the following
form:
\begin{eqnarray}
\label{res}
K^{(2)}=(\Phi\bar{\Phi})^2(c_1+c_2\ln\frac{\Phi\bar{\Phi}}{\mu_r}).
\end{eqnarray}
Here $c_1$ is a function of couplings involving the factor
$\frac{1}{\epsilon}$, $\epsilon=d-3$, where $d$ is a spacetime
dimension within the dimensional regularization prescription, and
$c_2$ is a function of couplings, which, however, is finite. The  
$\mu_r$ is a renormalization scale.

To obtain the renormalized effective K\"{a}hlerian potential, after
adding the corresponding counterterms, we impose the following
normalization condition:
\bea
\frac{\partial^2K}{\partial(\Phi\bar{\Phi})^2}|_{\Phi=\bar{\Phi}=v}=\lambda,
\eea
where $v$ is a mass scale, and $K=K^{(0)}+K^{(1)}+K^{(2)}$, where
$K^{(0)}=\frac{\lambda}{2}(\Phi\bar{\Phi})^2$, and 
$K^{(1)}$ and $K^{(2)}$ given by
eqs. (\ref{1loop}, \ref{res}) respectively, is the complete
K\"{a}hlerian effective potential up to two-loop order. Using this
condition to eliminate the dependence on $\mu_r$, we arrive at the
following renormalized K\"{a}hlerian effective potential:
\bea
\label{kren}
K_R=\frac{\lambda}{2}(\Phi\bar{\Phi})^2+
\frac{1}{8\pi}(5\lambda^2-\frac{1}{8}g^4)(\Phi\bar{\Phi})^2
-c_2(\Phi\bar{\Phi})^2[3-
\ln\frac{\Phi\bar{\Phi}}{v^2}].
\eea
We see that after the two-loop calculations, a mass scale has been
generated, breaking thus the scale invariance just as it occurs in the
Coleman-Weinberg model \cite{CW}.

In summary, in this paper we presented a superfield method for the
calculation of the effective potential in three-dimensional
supersymmetric gauge field theories. We succeeded to obtain explicit
expressions for the K\"ahlerian effective potential (which depends on
superfield $\Phi$ but not on its derivatives) up to two loops.  We
found that, in the two-loop order, a mass scale is generated, thus,
the scale invariance and superconformal symmetry are broken. 
One must emphasize the difference of our methodology from the one used in \cite{ourEP1}. While in that paper, the calculations were performed for a strongly restricted background field,  here we have done the calculations without any restriction on its structure, and  also did not use any component expansion. Nevertheless, the functional dependence of the effective action on the background fields is the same as in \cite{ourEP1} being given by (\ref{kren}). To carry out an exact calculation of the two-loop effective potential for this arbitrary background, however, one must calculate the contributions of the supergraphs depicted at Fig. 3, where the background dependent propagators are given by (\ref{vecpro}) and (\ref{scapro}). Such a calculation is extremely complicated from the technical viewpoint.
In principle, our approach can be directly generalized for higher loops,
and, also for theories with extended supersymmetry and
noncommutativity. Also, due to the similarity between the
structure of two-dimensional and three-dimensional supersymmetry
algebras, we expect that this approach can be also applied for the study
of the two-dimensional superfield theories.

\vspace{5mm}

{\bf Acknowledgments.}
This work was partially supported by Conselho Nacional de
Desenvolvimento Cient\'\i fico e Tecnol\'ogico (CNPq) and Funda\c
c\~ao de Apoio \`a
Pesquisa do  Estado de Rio Grande do Norte (FAPERN). A. Yu. P. has
been supported by the CNPq project No. 303461-2009/8. A. C. L. has
been supported by the CNPq project No.  303392/2010-0.

\end{document}